\documentclass[aps,prb,showpacs,floatfix,superscriptaddress,twocolumn,byrevtex]{revtex4-1}

\pdfoutput=1

\usepackage{amsmath}
\usepackage{amssymb}
\usepackage{amstext}
\usepackage{amsopn}
\usepackage{amsfonts}
\usepackage{amsxtra}

\usepackage{color}
\usepackage{dcolumn}
\usepackage{graphicx}
\usepackage{bm}

\usepackage{hyperref}

\newcommand{\degree}{\ensuremath{^\circ}}
\newcommand{\cm}{cm\ensuremath{^{-1}}}
\renewcommand{\d}[1]{\ensuremath{\operatorname{d}\!{#1}}}

\begin{document}

\title{Optical properties of BiTeBr and BiTeCl}

\author{Ana Akrap}
\email{ana.akrap@unige.ch}
\author{J\'er\'emie Teyssier}
\affiliation{DPMC, University of Geneva, CH-1211 Geneva 4, Switzerland}

\author{Arnaud Magrez}
\author{Philippe Bugnon}
\author{Helmuth Berger}
\affiliation{Crystal Growth Facility, Institut de Physique de la Mati\`ere Condens\'ee, \'Ecole Polytechnique F\'ed\'erale de Lausanne, Switzerland}

\author{Alexey B. Kuzmenko}
\author{Dirk van der Marel}
\affiliation{DPMC, University of Geneva, CH-1211 Geneva 4, Switzerland}

\date{\today}

\begin{abstract}
We present a comparative study of the optical properties -- reflectance, transmission and optical conductivity --  and Raman spectra of two layered bismuth-tellurohalides BiTeBr and BiTeCl at 300~K and 5~K, for light polarized in the {\em a-b} planes. Despite different space groups, the optical properties of the two compounds are very similar. Both materials are doped semiconductors, with the absorption edge above the optical gap which is lower in BiTeBr (0.62~eV) than in BiTeCl (0.77~eV). The same Rashba splitting is observed in the two materials. A non-Drude free carrier contribution in the optical conductivity, as well as three Raman and two infrared phonon modes, are observed in each compound. There is a dramatic difference in the highest infrared phonon intensity for the two compounds, and a difference in the doping levels. Aspects of the strong electron-phonon interaction are identified. Several interband transitions are assigned, among them the low-lying absorption $\beta$ which has the same value 0.25~eV in both compounds, and is caused by the Rashba spin splitting of the conduction band. An additional weak transition is found in BiTeCl, caused by the lower crystal symmetry.
\end{abstract}

%  PACS numbers
%  72.20.-i  Conductivity phenomena in semiconductors and insulators
%  78.20.-e  Optical properties of bulk materials and thin films
%  78.30.-j  Infrared and Raman spectra


\maketitle

\section{Introduction}
Large Rashba splitting of electron states arises in non-centrosymmetric structures with a strong spin-orbit coupling, and shifts bands of opposite spins away from each other in $k$-space. This effect manifests most clearly in two-dimensional structures, but also in certain bulk materials, namely semiconductors composed of heavy atoms. A series of bismuth tellurohalides BiTe$X$ ($X$=I, Br or Cl)  has recently been investigated from this perspective, and the possibility of their use as spintronics materials was raised.\cite{ishizaka11,crepaldi12} The effects of large Rashba spin splitting in BiTe$X$ were observed by ARPES,\cite{ishizaka11, landolt12, crepaldi12, sakano13} by optical spectroscopy in case of BiTeI,\cite{lee11} and is confirmed by several other experimental probes,\cite{demko12,tournier-colletta14} as well as band structure calculations.\cite{eremeev12}

Another aspect of the strong spin-orbit coupling in these systems which lack inversion symmetry is the possibility of stabilizing novel topological surface states. There were suggestions of the possible appearance of a topological insulator phase in BiTeI under high pressure,\cite{bahramy12, xi13} or the absence of this phase.\cite{tran14} More recently, assertions were made that topological surface states appear in BiTeCl at ambient pressure.\cite{chen13} While BiTeI has been investigated extensively, less is known about BiTeBr and BiTeCl. For example, the question of their band gap is unsettled.\cite{sakano13, chen13} The band structures of BiTeBr and BiTeCl are different from that of BiTeI,\cite{eremeev12,rusinov13} with a smaller Rashba splitting, which also raises the question of how this influences the interband electronic transitions and the signatures of the Rashba spin splitting.

In this paper, we address the optical properties of BiTeBr and BiTeCl at room temperature and at 5~K. We determine the respective absorption edges related to the optical gap, the positions and strengths of zone-center vibrational phonon modes, the spectral weight of the itinerant carriers, and characterize other interband transitions.  We find that the doping in these semiconductors is low, particularly in BiTeBr. The contribution of itinerant carriers has a non-Drude form which is possibly an indication of strong electron--phonon coupling, especially since one of the phonons in BiTeBr has a Fano shape. From the transmission measurement, we can assign the interband transitions between the Rashba-split bands. We find a qualitative difference between the two compounds by identifying a weak transition which is allowed in BiTeCl, but not in BiTeBr. Our data regarding interband transitions agree with the recent band structure calculations.\cite{rusinov13} 

\section{Experiment}
High quality single crystals of BiTeBr and BiTeCl are grown using two different techniques. BiTeBr crystals are obtained by chemical vapor transport from stoichiometric mixture of Bi, Te and BiBr$_3$, sealed with HBr as transport agent. The ampule is placed in a two-zone furnace with a charge and growth temperature of 440\degree C and 400\degree C, respectively. At the end of the growth processes, large centimeter-sized crystals are obtained. The structure and chemical composition are confirmed by X-ray diffraction and Energy Dispersive X-Ray Spectroscopy. BiTeCl single crystals are synthesized using a topotactic method, as previously described in Ref. \onlinecite{jacimovic14}. The samples of both compounds are kept in vacuum to preserve the stoichiometry.

\begin{figure}[!htb]
\centering
\includegraphics[width=0.9\linewidth]{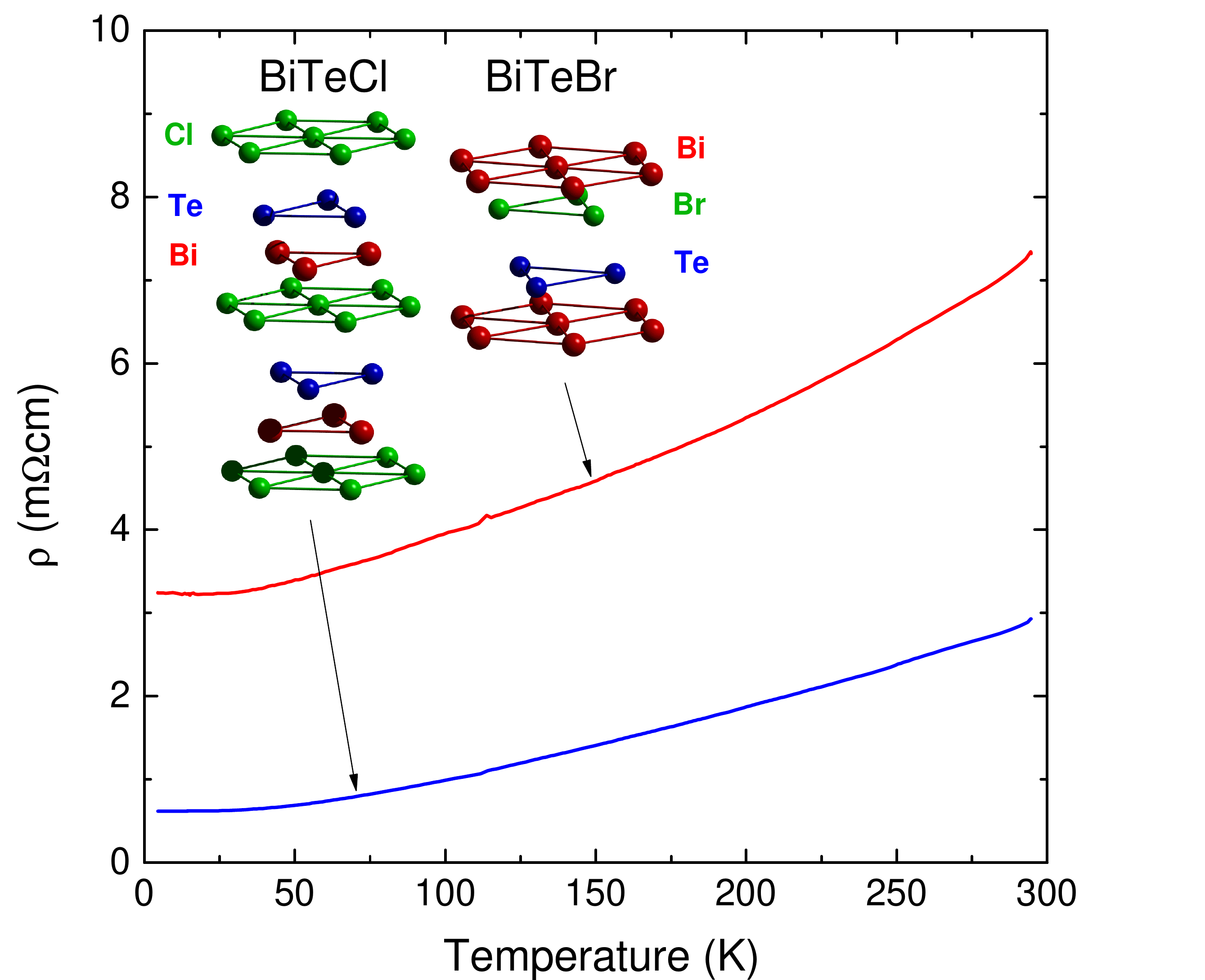}
\caption{The temperature dependence of the resistivity is shown for BiTeBr and BiTeCl, indicating metallic behavior. The crystal structures are also depicted, space group $P3m1$ for BiTeBr and $P6_3 mc$ for BiTeCl. Atoms of Bi, Te, and Br/Cl are shown in red, blue and green color, respectively.}
\label{fig:rho}
\end{figure}

Both structures are layered and non-centrosymmetric, containing a polar axis, and are shown as insets in Fig.~\ref{fig:rho}. They are slightly different: the space group for BiTeBr is the hexagonal $P3m1$, like in BiTeI, whereas for BiTeCl the stacking of Cl and Te atoms alternates along the $c$-axis resulting in a doubled unit cell and the $P6_3mc$ space group. Bismuth forms covalent bonds with Te and X atoms, and the interaction between Te and X atoms is much weaker.

The temperature-dependent reflectance is measured at a near-normal angle of incidence on freshly-cleaved surfaces for light polarized in the {\em a-b} planes, from $\simeq 15$ to $8\,000$~cm$^{-1}$ (2~meV to 1~eV) using {\em in situ} gold evaporation. Ellipsometry is used to determine the dielectric function from 4000 to 32000~\cm{} (0.5 to 4~eV). Transmission is measured on cleaved thin flakes using an infrared microscope in the midinfrared range, from 800 to 10000~\cm{} (0.1 -- 1.2~eV), with light polarized in the {\em a-b} planes. To obtain optical conductivity, we employ the Kramers-Kronig relation\cite{dressel-book} using suitable extrapolations for the reflectance in the $\omega \rightarrow 0,\infty$ limits, enhanced by the ellipsometric data which anchor the phase.\cite{kuzmenko05}

The Raman spectra are measured at room temperature without polarizers using a home-made micro Raman spectrometer equipped with an argon laser with a wavelength of 514.5~nm, a half-meter monochromator and a liquid nitrogen-cooled CCD detector. The spectral resolution is  $\sim 1$~cm$^{-1}$ and the laser power is kept low to prevent burning the sample surface. The resistivity is determined as a function of temperature using a four-point measurement.

%
% Figure 1
%

\section{Results and discussion}
Fig.~\ref{fig:rho} shows the resistivity $\rho(T)$ of BiTeBr and BiTeCl, which has a metallic temperature dependence in both compounds. The residual resistivity is higher in BiTeBr than in BiTeCl, suggesting a higher impurity concentration. Both systems are semiconductors, but are doped due to the off-stoichiometry which is mostly caused by a slight $X$ atom deficiency. The $n$-doping was found in the ARPES experiments\cite{sakano13} and it agrees with the negative thermopower.\cite{jacimovic14,kulbachinskii10} 

%
% Figure 2
%
\begin{figure}[htb]
\centering
\includegraphics[width=0.9\linewidth]{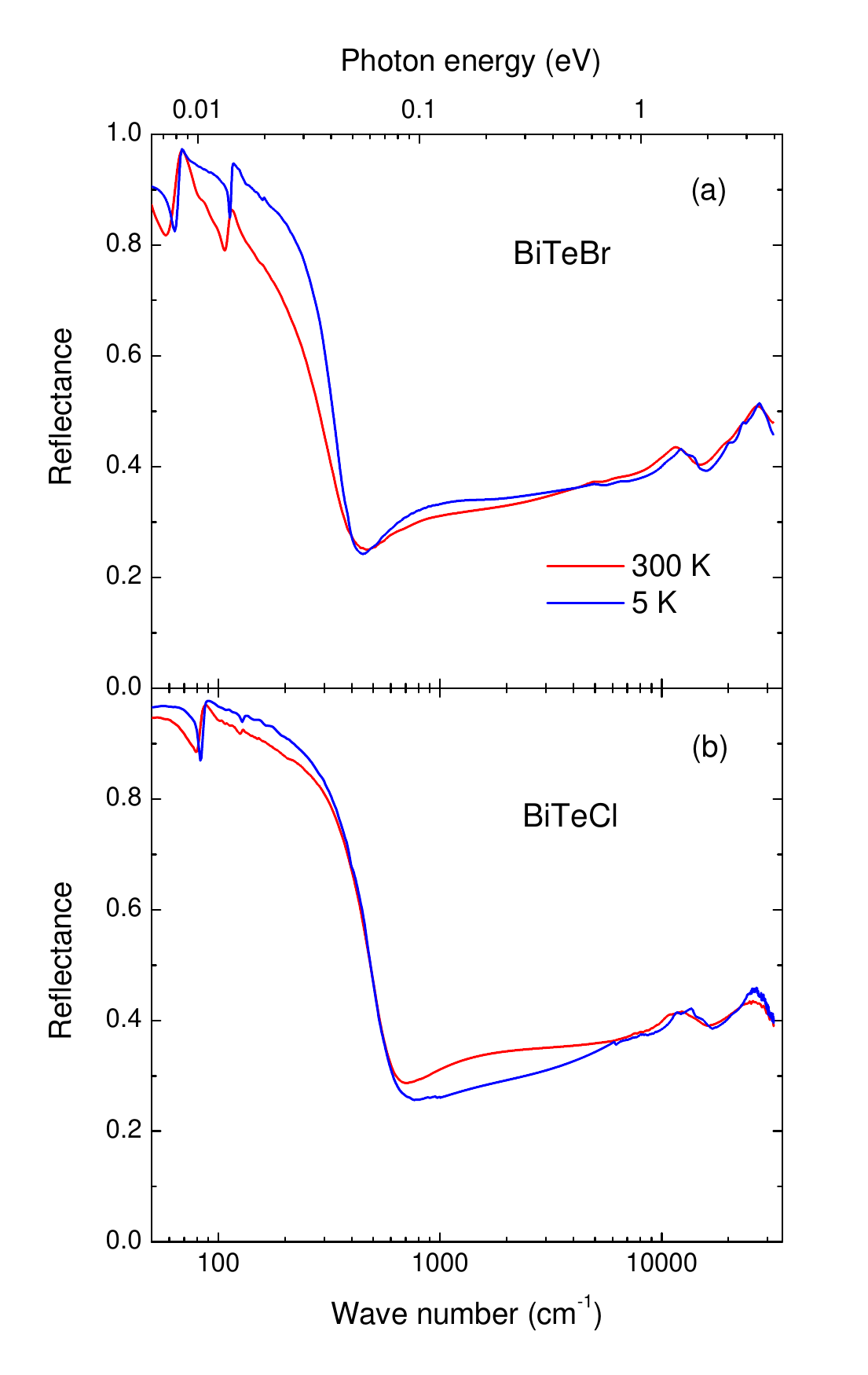}
\caption{Reflectance of (a) BiTeBr and (b) BiTeCl is shown in a wide frequency range in the $a$-$b$ plane.}
\label{fig:Refl}
\end{figure}

%
% Figure 3
%
\begin{figure}[htb]
\centering
\includegraphics[width=0.9\linewidth]{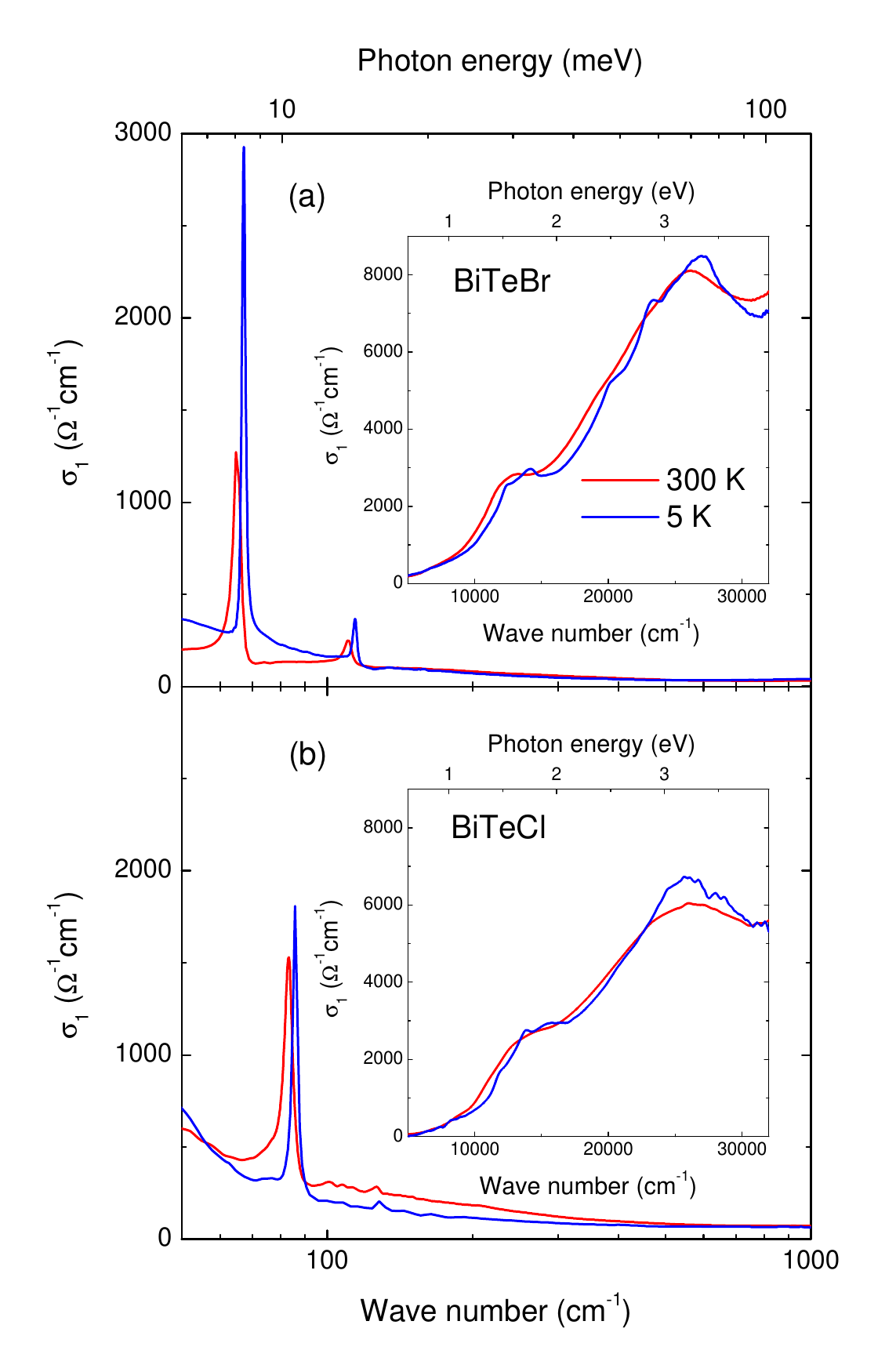}
\caption{The far-infrared range of $\sigma_1 (\omega)$ is shown in (a) for BiTeBr and (b) for BiTeCl. The insets show the midinfrared range $\sigma_1(\omega)$.}
\label{fig:sig1}
\end{figure}

The energy dependence of the reflectance is shown in Fig.~\ref{fig:Refl}(a) for BiTeBr and (b) for BiTeCl over a wide range. In both compounds, there is a pronounced plasma edge. In BiTeBr, the dip in the reflectance is at 400~\cm{}, compared to 600~\cm{} in BiTeCl. This difference is consistent with the higher conductivity of the studied BiTeCl samples. At energies below this edge the reflectance is high and approaches unity as $\omega \rightarrow 0$, signifying metallic conduction in agreement with transport measurements. The far-infrared sharp features in the reflectance can be attributed to the in-plane lattice vibrations. The high-energy peaks in the reflectance correspond to the interband transitions above the band gap.  In both compounds the reflectance at low energies increases as temperature is decreased, consistent with the metallic character, but the overall appearance does not change with temperature. 

If the reflectance is determined in a broad energy range, then the complex dielectric function $\hat{\epsilon}(\omega)$ can be obtained by a Kramers-Kronig transformation,\cite{dressel-book} and one can determine the optical conductivity:
$$\hat{\sigma}(\omega)=\sigma_1 +i\sigma_2 = \frac{i \omega[1-\hat{\epsilon}(\omega)]}{4\pi}$$
The real part of the optical conductivity, $\sigma_1(\omega)$, allows to clearly distinguish the itinerant carrier contribution, the interband transitions, as well as the vibrational phonon modes. In Fig.~\ref{fig:sig1} we show $\sigma_1(\omega)$ for BiTeBr  in panel (a), and for BiTeCl in (b).

%
% Table I - Drude Lorentz parameters
%
\begin{table*}[htb]
\caption{The bare plasma frequency $\omega_p(\omega)$, effective number of carriers $N_{eff}(\omega)$, and the phonon parameters (position $\omega$, width $\gamma_{ph}$ and phonon plasma frequency $\Omega$) of the Drude-Lorentz fits of the optical conductivity $\sigma_1(\omega)$ of BiTeBr and BiTeCl at $T=5$~K and 300~K are given below. Here $\omega_p(\omega)$ and $N_{eff}(\omega)$ are evaluated for $\omega=1000$~\cm{}. Positions of the Raman vibrational modes are given at 300~K, with mode symmetries indicated in parentheses. All frequencies are in cm$^{-1}$, and $N_{eff}$ is dimensionless.}
\vspace*{0.1cm}
\begin{ruledtabular}
\begin{tabular}{cc||cc|ccc|ccc|ccc}
 & & & &\multicolumn{3}{c|}{IR -- $E^1$}& \multicolumn{3}{c|}{IR -- $E^2$} & \multicolumn{3}{c}{Raman}\\
 &$T$(K) & $\omega_p$ & $N_{eff}$ & $\omega$ & $\gamma_{ph}$ & $\Omega$ & $\omega$ & $\gamma_{ph}$ & $\Omega$ & $\omega(E^1)$ & $\omega(E^2)$ & $\omega(A_1)$  \\
\cline{1-13}
BiTeBr & 5 & 1800 & 0.0037 & 67 & 2.0 & 520 & 114 & 2 & 182 &   &  &   \\
	    & 300 & 1660 & 0.0031 & 65 & 2.1 & 418 & 110 & 6 & 219 & $<62$ & 106 & 152 \\
\cline{1-13}
BiTeCl & 5 & 2460 & 0.0065 & 86 & 2 & 445  & 128 &3.6 & 124 &  &  & \\
	    & 300 & 2550 & 0.0070 & 83 & 3.3 & 493 & 126 &4.2 & 110 & 98 & 120 & 152 \\
\end{tabular}
\end{ruledtabular}
\label{tab:DLfit}
\end{table*}

The low energy $\sigma_1(\omega)$ shows a moderate Drude contribution, stronger in BiTeCl than in BiTeBr, and two phonon modes superimposed on the electronic background. Contrary to the free carrier contribution, the low-temperature infrared phonon modes are stronger in BiTeBr than in BiTeCl. The free carrier and bound excitations can be fit using a standard Drude-Lorentz model.\cite{dressel-book} However, the itinerant contribution in both compounds turns out to have a non-Drude shape and for a reliable fit two Drude-like contributions are needed. It is not clear if the non-Drude shape comes from the coexistence of two Fermi surfaces caused by Rashba splitting, or the complex form is due to a strong electron-phonon interaction. In the Drude-Lorentz fits, the second Drude component is characterized by a scattering time that is at least an order of magnitude larger than the main Drude component. To quantify the amount of itinerant carriers and their plasma frequency, we determine a frequency-dependent spectral weight. It is given by the area under the conductivity spectrum up to a chosen cutoff frequency $\omega$:
$$\frac{\omega_p^2(\omega)}{8}=\frac{\pi N_{eff}(\omega) e^2}{2m_e V_c}=\int_{0}^{\omega} \sigma_1(\omega') \d \omega',$$
where $\omega_p^2(\omega)$ is  the bare plasma frequency, $N_{eff}(\omega)$ is the effective number of carriers, $m_e$ is the free electron mass, and $V_c$ is the unit cell volume. Table~\ref{tab:DLfit} gives the values of $\omega_p^2$ and $N_{eff}$ at $\omega=1000$~\cm{}, a cutoff frequency for which the spectral weight contains most of the Drude peak, but which is below the interband transitions. All of the phonon contributions were subtracted from the spectral weight.  Bare plasma frequency $\omega_p$ at 5~K is 1800~\cm{} in BiTeBr, and 2460~\cm{} in BiTeCl. Because the effective mass $m^* < m_e$,\cite{KerrUnpublished} the shown $N_{eff}$ is an upper limit for the number of carriers per formula unit. In Table~\ref{tab:DLfit} one sees there are less than 0.3\% of electrons per formula unit in BiTeBr, and less than 0.7\% in BiTeCl. In BiTeCl $\omega_p$ and $N_{eff}$ are higher than in BiTeBr, which agrees with the higher doping and lower resistivity in BiTeCl. The Drude spectral weight does not significantly change with temperature.

Above 5000~\cm{}, $\sigma_1(\omega)$ is shown as an inset in Fig.~\ref{fig:sig1}(a) for BiTeBr, and (b) for BiTeCl. Overall, in this range, the optical conductivity is very similar for the two compounds. For both compounds it displays two pronounced peaks at 1.6 and 3.3~eV. Comparison with the band structure calculations\cite{eremeev12,rusinov13} suggests that the lower absorption (1.6-1.8~eV) corresponds to transitions between the bottom of the Rashba-split conduction band and the next higher band above the lowest conduction band at the $\Gamma$ point.

%
% Figure 4
%
\begin{figure}[htb]
\centering
\includegraphics[width=0.9\columnwidth]{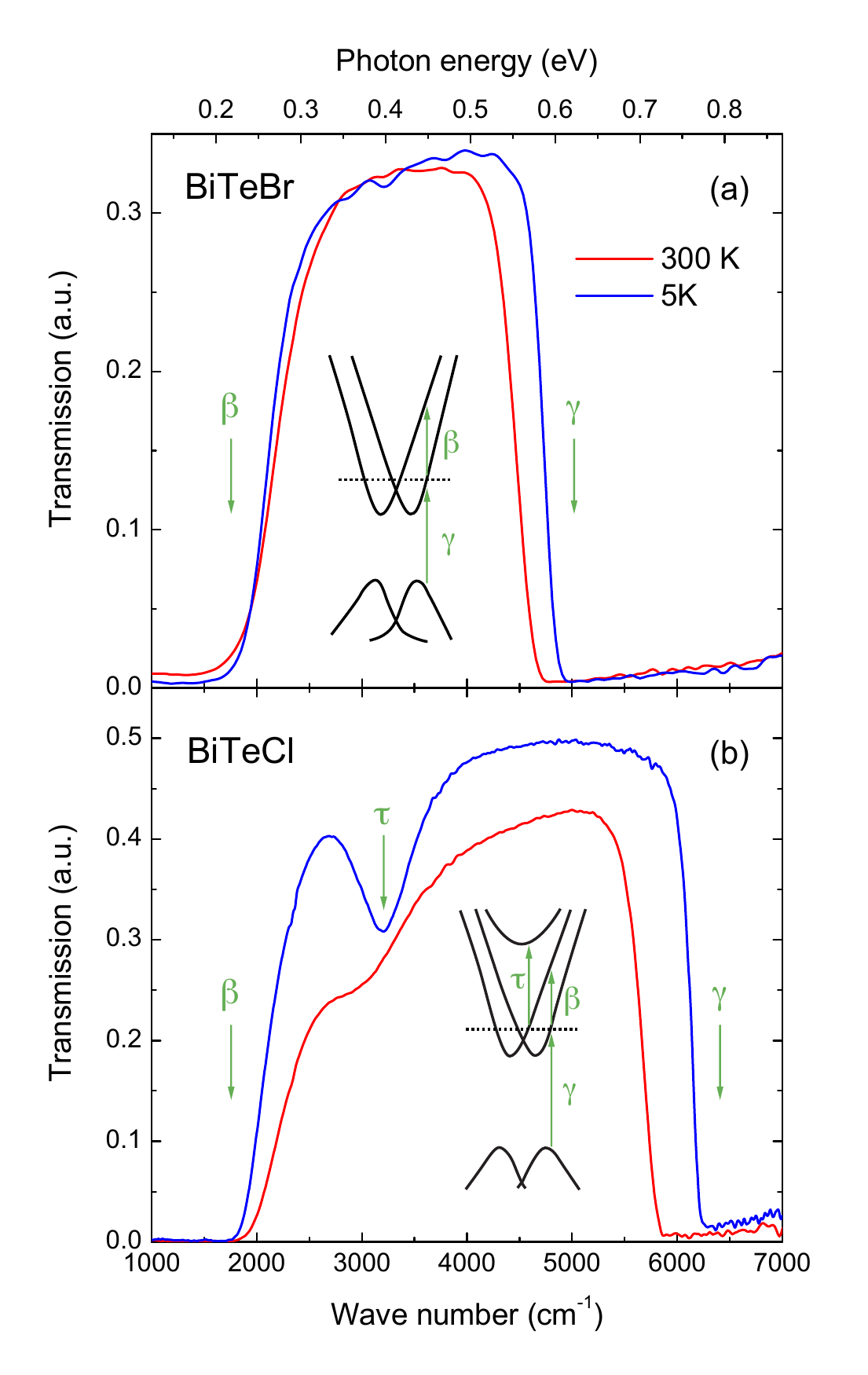}
\caption{Transmission through thin flakes of (a) BiTeBr and (b) BiTeCl shows a wide transparent range in the midinfrared. The insets show a sketch of the band structure for momentum parallel to the planes, adapted from Ref.~\onlinecite{rusinov13}.  The observed interband transitions are indicated: $\beta$, $\gamma$ and $\tau$.}
\label{fig:Trans}
\end{figure}

The spectral region in which $\sigma_1(\omega)$ is small, above the free carrier contribution and below strong mid-infrared absorptions, can be more precisely analyzed using transmission. Figure~\ref{fig:Trans}(a) and (b) shows transmission through thin flakes of BiTeBr (thickness 15~$\mu$m) and BiTeCl (thickness 10~$\mu$m), respectively, at 300~K and 5~K. In BiTeBr, the transmission is virtually nil below 2000~\cm{} and above 5000~\cm{}, and takes up finite values only in a narrow window between these two limits; a small transmission above 5000~\cm{} is an experimental artifact due to the detector nonlinearity. The low-energy edge of transmission corresponds to an interband transition. 
This edge is too sharp to be caused by a contribution from itinerant carriers, in addition it is temperature independent in contrast to the Drude absorption. We can assign this lower edge to the $\beta$ transition which connects the two Rashba-split electronic bands above the band gap, as illustrated in the band structure sketch in the inset of Fig.~\ref{fig:Trans}(a). 
The character of the low-lying conduction bands is Bi 6$p$, and the upper valence bands have predominantly Te 5$p$ character.\cite{sakano13}
The optical transitions between the two bands with different spin angular momenta are expected because of the spin-orbit interaction.\cite{sakano13}
The high-energy edge of transmission corresponds to the $\gamma$ transition, which is an excitation across the band gap. The absorption onset in BiTeBr occurs at 5000~\cm{} (0.62~eV) for $T=5$~K, and at a slightly lower frequency, 4750~\cm{} (0.59~eV) for 300~K. The room temperature value of the absorption edge is somewhat larger than in a previous optical absorption study,\cite{puga74} possibly due to a difference in sample quality. In BiTeCl, the transmission is similarly non-zero only in a well-defined window between 2000~\cm{} and 6000~\cm{} at 300~K. Similarly, the lower limit may be attributed to the $\beta$ absorption, and the higher limit corresponds to the onset of absorption, $\gamma$, equal to 6250~\cm{} (0.77~eV) at 5~K and 5800~\cm{} (0.72~eV) at 300~K. 
% Here add why beta can be observed between states of different spin.

The Cl ion is more electronegative than Br or I, so that the states below the gap in BiTeCl are pulled lower in energy with respect to BiTeBr and BiTeI,\cite{eremeev12,rusinov13} and the absorption edge $\gamma$ in BiTeCl is higher than in BiTeBr. For comparison, in BiTeI  this energy is 0.4~eV.\cite{lee11,tran14} However, not only does the gap vary through this series, but also the position of the chemical potential is sample-dependent. As mentioned above, for the samples used in this study, the chemical potential is higher in the BiTeCl samples than in BiTeBr. This may in part be responsible for a further small shift of the absorption edge in BiTeCl to higher values. Our values of $\gamma$ compare well to the LDA+GW calculation of band gaps\cite{rusinov13} that take into account the interactions, and give a gap of 0.65~eV in BiTeBr and 0.87~eV in BiTeCl. Unlike the $\gamma$ transition, the $\beta$ transition is almost at the same frequency $\sim 2000$~\cm{} (0.25~eV) in BiTeBr and BiTeCl, and it shows very little temperature dependence. For comparison, $\beta$ was found to be at a higher energy $\sim 0.35$~eV in BiTeI,\cite{lee11} consistent with the stronger Rashba spin splitting in that compound caused by the larger mass of iodine compared to Br and Cl.\cite{eremeev12}

In addition to the two prominent interband transitions $\beta$ and $\gamma$, there is a distinct - albeit much weaker - feature in BiTeCl, here referred to as $\tau$, which manifests as a small dip in the transmission spectra. This absorption is situated close to 3000~\cm{} at 300~K, and hardens to 3200~\cm{} (0.4~eV) at 5~K while becoming more pronounced. It is likely the result of a band transition which is  weakly allowed in BiTeCl, but is forbidden in BiTeBr. The BiTeCl structure has a lower symmetry and its unit cell is doubled along the $c$-axis. When the bands are folded into the new Brillouin zone, a new set of bands appears above the lowest bismuth bands in BiTeCl.\cite{eremeev12} Indeed, a suitable candidate for such an interband transition below the band gap can be distinguished in the calculated band structures\cite{rusinov13} at $\sim 0.8$~eV above the top of the highest valence band at $\Gamma$ point, as illustrated in the inset of Fig.~\ref{fig:Trans}(b).

%
% Figure 5
%
\begin{figure}[!htb]
\centering
\includegraphics[width=0.9\columnwidth]{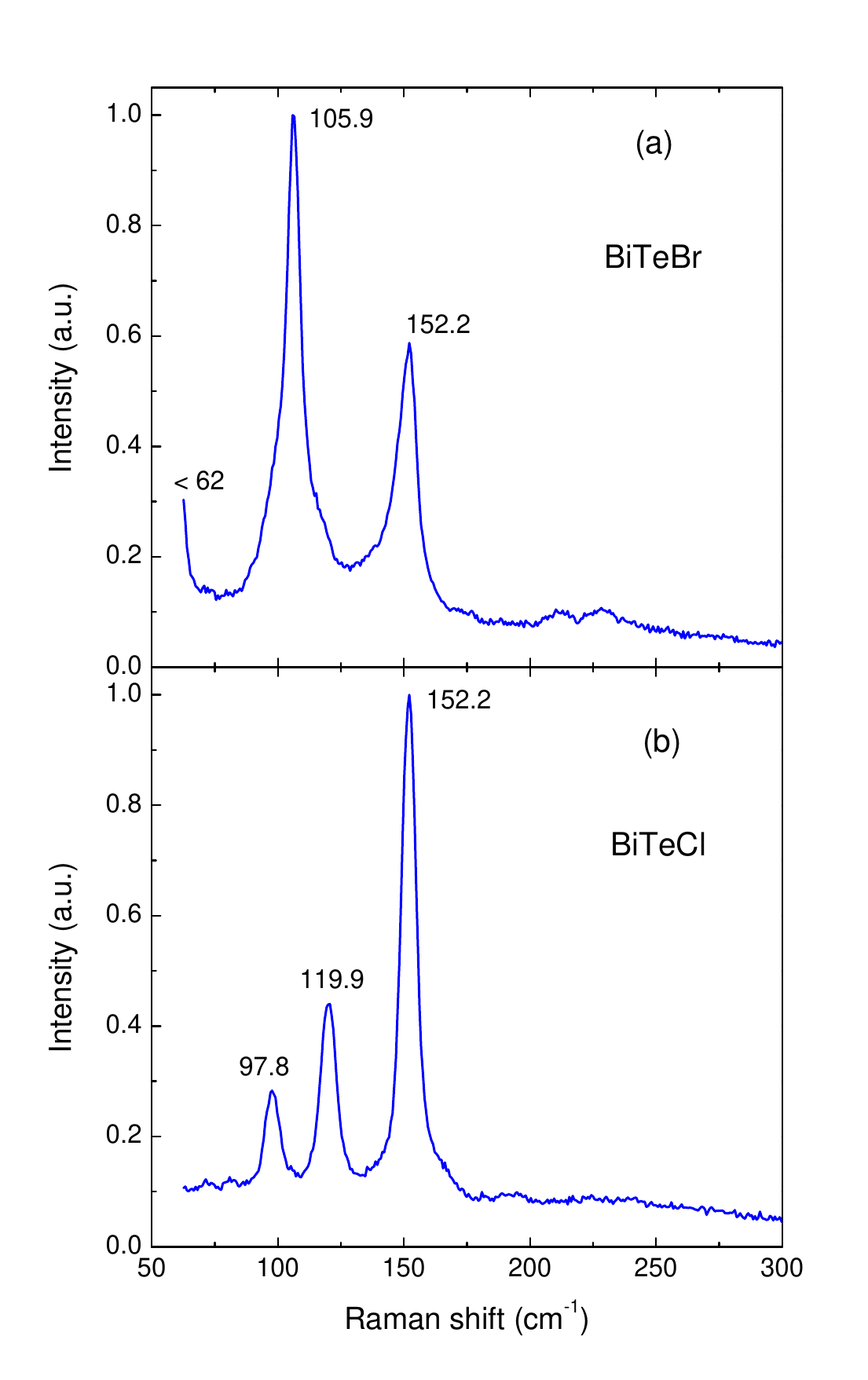}
\caption{Raman spectra for  (a) BiTeBr and (b) BiTeCl taken at 300~K show three Raman active modes in each compound.}
\label{fig:Raman}
\end{figure}

Finally, we discuss the vibrational phonon modes in these two systems. BiTeBr, like BiTeI, is characterized by four zone-center vibrational modes: $\Gamma=2A_1+2E$. Both the $A_1$ and the doubly degenerate $E$ modes are simultaneously Raman and infrared active due to the lack of inversion symmetry. While the $A_1$ modes are infrared active only along the polar $c$ axis, and are therefore not observable, the $E$ modes are active in the \emph{a-b} planes and can be detected in our experimental configuration. This is consistent with the observation of two infrared-active modes, as shown in Fig.~\ref{fig:sig1}(a). For BiTeCl the irreducible vibrational representation is more complex:  $\Gamma = 2A_1 + 3B_1 + 2E_1 + 3E_2$. Similar to BiTeBr, the only in-plane infrared-active modes are the two $E_1$ modes, which are indeed observed (Fig.~\ref{fig:sig1}(b)), whereas the two out-of-plane $A_1$ modes cannot be detected with light polarized in-plane. Apart from the silent $B_1$ modes, all other vibrational modes are Raman active.

Fig.~\ref{fig:Raman} shows the room-temperature Raman spectra in both compounds. For BiTeBr one can identify two strong vibrational modes, along with one weaker mode, and an onset of a low-frequency mode: 105.9, 152.2 \cm{}, a weak shoulder at 115~\cm{}, and a mode below 62 \cm{}, which is the cutoff of our experimental setup. The positions of the peaks are somewhat different than in the previously reported Raman spectra of BiTeBr, where Sklyadneva \emph{et al.}\cite{sklyadneva12} observe three modes at 58, 98.5 and 130.5~\cm{}, all shifted to lower energies with respect to our data. The reason for this discrepancy is unclear. On the contrary, the positions of Raman active modes in BiTeCl are very similar to the previously determined frequencies.\cite{sklyadneva12} As indicated in Fig.~\ref{fig:Raman}, three modes are observed in BiTeCl at 97.8, 119.9 and 152.2~\cm{}. 

The phonon frequencies determined from the Raman spectra at 300~K are detailed in Table~\ref{tab:DLfit}, along with their symmetries. Also listed  are the Lorentzian fit parameters for the two resolved infrared phonons at 5 and 300~K: the vibration frequency $\omega$, the width $\gamma_{ph}$ and the plasma frequency $\Omega$. The two $E$ ($E_1$ in BiTeCl) modes were assigned by Sklyadneva \emph{et al,}\cite{sklyadneva12} specifying that while both $E$-type modes involve in-plane vibration of the Bi, Te and $X$ layers, the lower $E$ mode has a large contribution of the $X$ atom vibration. Hence this mode blueshifts significantly when replacing Br by the lighter Cl; the mode shifts from 67~\cm{} at 5~K in BiTeBr to 85~\cm{} in BiTeCl. The highest observed mode is of $A_1$ symmetry,\cite{sklyadneva12} and is produced by a motion of Bi and Te sublattices against each other along the polar axis. This mode is independent of the halogen atom and is at the same frequency in BiTeBr and BiTeCl, 152~\cm{}, very close to the high frequency $A_1$ mode in BiTeI which is observed at 150~\cm{}.\cite{tran14}

The fact that infrared and Raman modes are observed at similar frequencies is consistent with the absence of an inversion center. In BiTeBr, the higher $E$ mode appears at 110~\cm{} in the infrared, and at 106~\cm{} in the Raman spectrum. Similarly, in BiTeCl the higher $E_1$ mode is at 126~\cm{} in the infrared, and at 120~\cm{} in the Raman spectrum. There is, however, a discrepancy in the lower $E_1$ mode in BiTeCl, which is at 82~\cm{} in the infrared, and at 98~\cm{} in the Raman spectrum. 

All of the infrared vibrational modes can be well fit by the standard Drude-Lorentz model. However, fitting a Fano asymmetric line shape\cite{damascelli97} considerably improves the quality of the fit for the higher frequency $E$ mode in BiTeBr, and yields a significant Fano parameter $q \sim 2.9$ at 5~K (for $q \rightarrow \infty$ the mode would be perfectly symmetric, with a Lorentzian lineshape). The corresponding $E_1$ mode in BiTeCl does not show clear Fano asymmetry. As stated earlier, the non-Drude shape of the free carrier contribution may be related to a strong electron-phonon interaction, poorly screened because there are few free carriers. This interaction can result in Fano shapes of the phonon vibrations.  

\section{Conclusions}
We identify the interband transition $\beta$ at 0.25~eV within both compounds, which implies that the Rashba splitting is the same. In BiTeCl a weak transition $\tau$ is found at 0.4~eV, linked to a lower crystal symmetry. The low temperature optical absorption edge $\gamma$ is at 0.62~eV in BiTeBr, and 0.77~eV in BiTeCl. From the Drude spectral weight we deduce that both systems have rather low doping. The high energy transitions agree well with the calculated band structure.\cite{eremeev12,rusinov13} Three Raman active modes are identified in each system, and two normally-active $E$-type phonon modes are observed in the infrared spectra. An asymmetric Fano shape in the higher $E$ phonon and the non-Drude form of the free carrier optical conductivity indicate that the electron-phonon coupling may be important in these compounds.

\begin{acknowledgements}
Research was supported by the Swiss NSF through grant No. 200020-135085 and its NCCR MaNEP. A.A. acknowledges funding from `` The Excellence Fellowship'' of the University of Geneva.
\end{acknowledgements}
\vfil

\bibliography{BiTeX}

\providecommand{\noopsort}[1]{}\providecommand{\singleletter}[1]{#1}
\begin{thebibliography}{21}%
\makeatletter
\providecommand \@ifxundefined [1]{%
 \@ifx{#1\undefined}
}%
\providecommand \@ifnum [1]{%
 \ifnum #1\expandafter \@firstoftwo
 \else \expandafter \@secondoftwo
 \fi
}%
\providecommand \@ifx [1]{%
 \ifx #1\expandafter \@firstoftwo
 \else \expandafter \@secondoftwo
 \fi
}%
\providecommand \natexlab [1]{#1}%
\providecommand \enquote  [1]{``#1''}%
\providecommand \bibnamefont  [1]{#1}%
\providecommand \bibfnamefont [1]{#1}%
\providecommand \citenamefont [1]{#1}%
\providecommand \href@noop [0]{\@secondoftwo}%
\providecommand \href [0]{\begingroup \@sanitize@url \@href}%
\providecommand \@href[1]{\@@startlink{#1}\@@href}%
\providecommand \@@href[1]{\endgroup#1\@@endlink}%
\providecommand \@sanitize@url [0]{\catcode `\\12\catcode `\$12\catcode
  `\&12\catcode `\#12\catcode `\^12\catcode `\_12\catcode `\%12\relax}%
\providecommand \@@startlink[1]{}%
\providecommand \@@endlink[0]{}%
\providecommand \url  [0]{\begingroup\@sanitize@url \@url }%
\providecommand \@url [1]{\endgroup\@href {#1}{\urlprefix }}%
\providecommand \urlprefix  [0]{URL }%
\providecommand \Eprint [0]{\href }%
\providecommand \doibase [0]{http://dx.doi.org/}%
\providecommand \selectlanguage [0]{\@gobble}%
\providecommand \bibinfo  [0]{\@secondoftwo}%
\providecommand \bibfield  [0]{\@secondoftwo}%
\providecommand \translation [1]{[#1]}%
\providecommand \BibitemOpen [0]{}%
\providecommand \bibitemStop [0]{}%
\providecommand \bibitemNoStop [0]{.\EOS\space}%
\providecommand \EOS [0]{\spacefactor3000\relax}%
\providecommand \BibitemShut  [1]{\csname bibitem#1\endcsname}%
\let\auto@bib@innerbib\@empty
%</preamble>
\bibitem [{\citenamefont {Ishizaka}\ \emph {et~al.}(2011)\citenamefont
  {Ishizaka}, \citenamefont {Bahramy}, \citenamefont {Murakawa}, \citenamefont
  {Sakano}, \citenamefont {Shimojima}, \citenamefont {Sonobe}, \citenamefont
  {Koizumi}, \citenamefont {Shin}, \citenamefont {Miyahara}, \citenamefont
  {Kimura}, \citenamefont {Miyamoto}, \citenamefont {Okuda}, \citenamefont
  {Namatame}, \citenamefont {Taniguchi}, \citenamefont {Arita}, \citenamefont
  {Nagaosa}, \citenamefont {Kobayashi}, \citenamefont {Murakami}, \citenamefont
  {Kumai}, \citenamefont {Kaneko}, \citenamefont {Onose},\ and\ \citenamefont
  {Tokura}}]{ishizaka11}%
  \BibitemOpen
  \bibfield  {author} {\bibinfo {author} {\bibfnamefont {K.}~\bibnamefont
  {Ishizaka}}, \bibinfo {author} {\bibfnamefont {M.~S.}\ \bibnamefont
  {Bahramy}}, \bibinfo {author} {\bibfnamefont {H.}~\bibnamefont {Murakawa}},
  \bibinfo {author} {\bibfnamefont {M.}~\bibnamefont {Sakano}}, \bibinfo
  {author} {\bibfnamefont {T.}~\bibnamefont {Shimojima}}, \bibinfo {author}
  {\bibfnamefont {T.}~\bibnamefont {Sonobe}}, \bibinfo {author} {\bibfnamefont
  {K.}~\bibnamefont {Koizumi}}, \bibinfo {author} {\bibfnamefont
  {S.}~\bibnamefont {Shin}}, \bibinfo {author} {\bibfnamefont {H.}~\bibnamefont
  {Miyahara}}, \bibinfo {author} {\bibfnamefont {A.}~\bibnamefont {Kimura}},
  \bibinfo {author} {\bibfnamefont {K.}~\bibnamefont {Miyamoto}}, \bibinfo
  {author} {\bibfnamefont {T.}~\bibnamefont {Okuda}}, \bibinfo {author}
  {\bibfnamefont {H.}~\bibnamefont {Namatame}}, \bibinfo {author}
  {\bibfnamefont {M.}~\bibnamefont {Taniguchi}}, \bibinfo {author}
  {\bibfnamefont {R.}~\bibnamefont {Arita}}, \bibinfo {author} {\bibfnamefont
  {N.}~\bibnamefont {Nagaosa}}, \bibinfo {author} {\bibfnamefont
  {K.}~\bibnamefont {Kobayashi}}, \bibinfo {author} {\bibfnamefont
  {Y.}~\bibnamefont {Murakami}}, \bibinfo {author} {\bibfnamefont
  {R.}~\bibnamefont {Kumai}}, \bibinfo {author} {\bibfnamefont
  {Y.}~\bibnamefont {Kaneko}}, \bibinfo {author} {\bibfnamefont
  {Y.}~\bibnamefont {Onose}}, \ and\ \bibinfo {author} {\bibfnamefont
  {Y.}~\bibnamefont {Tokura}},\ }\href {\doibase 10.1038/nmat3051} {\bibfield
  {journal} {\bibinfo  {journal} {Nature Materials}\ }\textbf {\bibinfo
  {volume} {10}},\ \bibinfo {pages} {521} (\bibinfo {year} {2011})}\BibitemShut
  {NoStop}%
\bibitem [{\citenamefont {Crepaldi}\ \emph {et~al.}(2012)\citenamefont
  {Crepaldi}, \citenamefont {Moreschini}, \citenamefont {Aut{\`e}s},
  \citenamefont {Tournier-Colletta}, \citenamefont {Moser}, \citenamefont
  {Virk}, \citenamefont {Berger}, \citenamefont {Bugnon}, \citenamefont
  {Chang}, \citenamefont {Kern}, \citenamefont {Bostwick}, \citenamefont
  {Rotenberg}, \citenamefont {Yazyev},\ and\ \citenamefont
  {Grioni}}]{crepaldi12}%
  \BibitemOpen
  \bibfield  {author} {\bibinfo {author} {\bibfnamefont {A.}~\bibnamefont
  {Crepaldi}}, \bibinfo {author} {\bibfnamefont {L.}~\bibnamefont
  {Moreschini}}, \bibinfo {author} {\bibfnamefont {G.}~\bibnamefont
  {Aut{\`e}s}}, \bibinfo {author} {\bibfnamefont {C.}~\bibnamefont
  {Tournier-Colletta}}, \bibinfo {author} {\bibfnamefont {S.}~\bibnamefont
  {Moser}}, \bibinfo {author} {\bibfnamefont {N.}~\bibnamefont {Virk}},
  \bibinfo {author} {\bibfnamefont {H.}~\bibnamefont {Berger}}, \bibinfo
  {author} {\bibfnamefont {P.}~\bibnamefont {Bugnon}}, \bibinfo {author}
  {\bibfnamefont {Y.~J.}\ \bibnamefont {Chang}}, \bibinfo {author}
  {\bibfnamefont {K.}~\bibnamefont {Kern}}, \bibinfo {author} {\bibfnamefont
  {A.}~\bibnamefont {Bostwick}}, \bibinfo {author} {\bibfnamefont
  {E.}~\bibnamefont {Rotenberg}}, \bibinfo {author} {\bibfnamefont {O.~V.}\
  \bibnamefont {Yazyev}}, \ and\ \bibinfo {author} {\bibfnamefont
  {M.}~\bibnamefont {Grioni}},\ }\href
  {http://link.aps.org/doi/10.1103/PhysRevLett.109.096803} {\bibfield
  {journal} {\bibinfo  {journal} {Physical Review Letters}\ }\textbf {\bibinfo
  {volume} {109}},\ \bibinfo {pages} {096803} (\bibinfo {year}
  {2012})}\BibitemShut {NoStop}%
\bibitem [{\citenamefont {Landolt}\ \emph {et~al.}(2012)\citenamefont
  {Landolt}, \citenamefont {Eremeev}, \citenamefont {Koroteev}, \citenamefont
  {Slomski}, \citenamefont {Muff}, \citenamefont {Neupert}, \citenamefont
  {Kobayashi}, \citenamefont {Strocov}, \citenamefont {Schmitt}, \citenamefont
  {Aliev}, \citenamefont {Babanly}, \citenamefont {Amiraslanov}, \citenamefont
  {Chulkov}, \citenamefont {Osterwalder},\ and\ \citenamefont
  {Dil}}]{landolt12}%
  \BibitemOpen
  \bibfield  {author} {\bibinfo {author} {\bibfnamefont {G.}~\bibnamefont
  {Landolt}}, \bibinfo {author} {\bibfnamefont {S.~V.}\ \bibnamefont
  {Eremeev}}, \bibinfo {author} {\bibfnamefont {Y.~M.}\ \bibnamefont
  {Koroteev}}, \bibinfo {author} {\bibfnamefont {B.}~\bibnamefont {Slomski}},
  \bibinfo {author} {\bibfnamefont {S.}~\bibnamefont {Muff}}, \bibinfo {author}
  {\bibfnamefont {T.}~\bibnamefont {Neupert}}, \bibinfo {author} {\bibfnamefont
  {M.}~\bibnamefont {Kobayashi}}, \bibinfo {author} {\bibfnamefont {V.~N.}\
  \bibnamefont {Strocov}}, \bibinfo {author} {\bibfnamefont {T.}~\bibnamefont
  {Schmitt}}, \bibinfo {author} {\bibfnamefont {Z.~S.}\ \bibnamefont {Aliev}},
  \bibinfo {author} {\bibfnamefont {M.~B.}\ \bibnamefont {Babanly}}, \bibinfo
  {author} {\bibfnamefont {I.~R.}\ \bibnamefont {Amiraslanov}}, \bibinfo
  {author} {\bibfnamefont {E.~V.}\ \bibnamefont {Chulkov}}, \bibinfo {author}
  {\bibfnamefont {J.}~\bibnamefont {Osterwalder}}, \ and\ \bibinfo {author}
  {\bibfnamefont {J.~H.}\ \bibnamefont {Dil}},\ }\href
  {http://link.aps.org/doi/10.1103/PhysRevLett.109.116403} {\bibfield
  {journal} {\bibinfo  {journal} {Physical Review Letters}\ }\textbf {\bibinfo
  {volume} {109}},\ \bibinfo {pages} {116403} (\bibinfo {year}
  {2012})}\BibitemShut {NoStop}%
\bibitem [{\citenamefont {Sakano}\ \emph {et~al.}(2013)\citenamefont {Sakano},
  \citenamefont {Bahramy}, \citenamefont {Katayama}, \citenamefont {Shimojima},
  \citenamefont {Murakawa}, \citenamefont {Kaneko}, \citenamefont {Malaeb},
  \citenamefont {Shin}, \citenamefont {Ono}, \citenamefont {Kumigashira},
  \citenamefont {Arita}, \citenamefont {Nagaosa}, \citenamefont {Hwang},
  \citenamefont {Tokura},\ and\ \citenamefont {Ishizaka}}]{sakano13}%
  \BibitemOpen
  \bibfield  {author} {\bibinfo {author} {\bibfnamefont {M.}~\bibnamefont
  {Sakano}}, \bibinfo {author} {\bibfnamefont {M.~S.}\ \bibnamefont {Bahramy}},
  \bibinfo {author} {\bibfnamefont {A.}~\bibnamefont {Katayama}}, \bibinfo
  {author} {\bibfnamefont {T.}~\bibnamefont {Shimojima}}, \bibinfo {author}
  {\bibfnamefont {H.}~\bibnamefont {Murakawa}}, \bibinfo {author}
  {\bibfnamefont {Y.}~\bibnamefont {Kaneko}}, \bibinfo {author} {\bibfnamefont
  {W.}~\bibnamefont {Malaeb}}, \bibinfo {author} {\bibfnamefont
  {S.}~\bibnamefont {Shin}}, \bibinfo {author} {\bibfnamefont {K.}~\bibnamefont
  {Ono}}, \bibinfo {author} {\bibfnamefont {H.}~\bibnamefont {Kumigashira}},
  \bibinfo {author} {\bibfnamefont {R.}~\bibnamefont {Arita}}, \bibinfo
  {author} {\bibfnamefont {N.}~\bibnamefont {Nagaosa}}, \bibinfo {author}
  {\bibfnamefont {H.~Y.}\ \bibnamefont {Hwang}}, \bibinfo {author}
  {\bibfnamefont {Y.}~\bibnamefont {Tokura}}, \ and\ \bibinfo {author}
  {\bibfnamefont {K.}~\bibnamefont {Ishizaka}},\ }\href
  {http://link.aps.org/doi/10.1103/PhysRevLett.110.107204} {\bibfield
  {journal} {\bibinfo  {journal} {Physical Review Letters}\ }\textbf {\bibinfo
  {volume} {110}},\ \bibinfo {pages} {107204} (\bibinfo {year}
  {2013})}\BibitemShut {NoStop}%
\bibitem [{\citenamefont {Lee}\ \emph {et~al.}(2011)\citenamefont {Lee},
  \citenamefont {Schober}, \citenamefont {Bahramy}, \citenamefont {Murakawa},
  \citenamefont {Onose}, \citenamefont {Arita}, \citenamefont {Nagaosa},\ and\
  \citenamefont {Tokura}}]{lee11}%
  \BibitemOpen
  \bibfield  {author} {\bibinfo {author} {\bibfnamefont {J.~S.}\ \bibnamefont
  {Lee}}, \bibinfo {author} {\bibfnamefont {G.~A.~H.}\ \bibnamefont {Schober}},
  \bibinfo {author} {\bibfnamefont {M.~S.}\ \bibnamefont {Bahramy}}, \bibinfo
  {author} {\bibfnamefont {H.}~\bibnamefont {Murakawa}}, \bibinfo {author}
  {\bibfnamefont {Y.}~\bibnamefont {Onose}}, \bibinfo {author} {\bibfnamefont
  {R.}~\bibnamefont {Arita}}, \bibinfo {author} {\bibfnamefont
  {N.}~\bibnamefont {Nagaosa}}, \ and\ \bibinfo {author} {\bibfnamefont
  {Y.}~\bibnamefont {Tokura}},\ }\href@noop {} {\bibfield  {journal} {\bibinfo
  {journal} {Physical Review Letters}\ }\textbf {\bibinfo {volume} {107}},\
  \bibinfo {pages} {117401} (\bibinfo {year} {2011})}\BibitemShut {NoStop}%
\bibitem [{\citenamefont {Demk{\'o}}\ \emph {et~al.}(2012)\citenamefont
  {Demk{\'o}}, \citenamefont {Schober}, \citenamefont {Kocsis}, \citenamefont
  {Bahramy}, \citenamefont {Murakawa}, \citenamefont {Lee}, \citenamefont
  {K{\'e}zsm{\'a}rki}, \citenamefont {Arita}, \citenamefont {Nagaosa},\ and\
  \citenamefont {Tokura}}]{demko12}%
  \BibitemOpen
  \bibfield  {author} {\bibinfo {author} {\bibfnamefont {L.}~\bibnamefont
  {Demk{\'o}}}, \bibinfo {author} {\bibfnamefont {G.~A.~H.}\ \bibnamefont
  {Schober}}, \bibinfo {author} {\bibfnamefont {V.}~\bibnamefont {Kocsis}},
  \bibinfo {author} {\bibfnamefont {M.~S.}\ \bibnamefont {Bahramy}}, \bibinfo
  {author} {\bibfnamefont {H.}~\bibnamefont {Murakawa}}, \bibinfo {author}
  {\bibfnamefont {J.~S.}\ \bibnamefont {Lee}}, \bibinfo {author} {\bibfnamefont
  {I.}~\bibnamefont {K{\'e}zsm{\'a}rki}}, \bibinfo {author} {\bibfnamefont
  {R.}~\bibnamefont {Arita}}, \bibinfo {author} {\bibfnamefont
  {N.}~\bibnamefont {Nagaosa}}, \ and\ \bibinfo {author} {\bibfnamefont
  {Y.}~\bibnamefont {Tokura}},\ }\href
  {http://link.aps.org/doi/10.1103/PhysRevLett.109.167401} {\bibfield
  {journal} {\bibinfo  {journal} {Physical Review Letters}\ }\textbf {\bibinfo
  {volume} {109}},\ \bibinfo {pages} {167401} (\bibinfo {year}
  {2012})}\BibitemShut {NoStop}%
\bibitem [{\citenamefont {Tournier-Colletta}\ \emph {et~al.}(2014)\citenamefont
  {Tournier-Colletta}, \citenamefont {Aut{\`e}s}, \citenamefont {Kierren},
  \citenamefont {Bugnon}, \citenamefont {Berger}, \citenamefont
  {Fagot-Revurat}, \citenamefont {Yazyev}, \citenamefont {Grioni},\ and\
  \citenamefont {Malterre}}]{tournier-colletta14}%
  \BibitemOpen
  \bibfield  {author} {\bibinfo {author} {\bibfnamefont {C.}~\bibnamefont
  {Tournier-Colletta}}, \bibinfo {author} {\bibfnamefont {G.}~\bibnamefont
  {Aut{\`e}s}}, \bibinfo {author} {\bibfnamefont {B.}~\bibnamefont {Kierren}},
  \bibinfo {author} {\bibfnamefont {P.}~\bibnamefont {Bugnon}}, \bibinfo
  {author} {\bibfnamefont {H.}~\bibnamefont {Berger}}, \bibinfo {author}
  {\bibfnamefont {Y.}~\bibnamefont {Fagot-Revurat}}, \bibinfo {author}
  {\bibfnamefont {O.~V.}\ \bibnamefont {Yazyev}}, \bibinfo {author}
  {\bibfnamefont {M.}~\bibnamefont {Grioni}}, \ and\ \bibinfo {author}
  {\bibfnamefont {D.}~\bibnamefont {Malterre}},\ }\href
  {http://link.aps.org/doi/10.1103/PhysRevB.89.085402} {\bibfield  {journal}
  {\bibinfo  {journal} {Physical Review B}\ }\textbf {\bibinfo {volume} {89}},\
  \bibinfo {pages} {085402} (\bibinfo {year} {2014})}\BibitemShut {NoStop}%
\bibitem [{\citenamefont {Eremeev}\ \emph {et~al.}(2012)\citenamefont
  {Eremeev}, \citenamefont {Nechaev}, \citenamefont {Koroteev}, \citenamefont
  {Echenique},\ and\ \citenamefont {Chulkov}}]{eremeev12}%
  \BibitemOpen
  \bibfield  {author} {\bibinfo {author} {\bibfnamefont {S.~V.}\ \bibnamefont
  {Eremeev}}, \bibinfo {author} {\bibfnamefont {I.~A.}\ \bibnamefont
  {Nechaev}}, \bibinfo {author} {\bibfnamefont {Y.~M.}\ \bibnamefont
  {Koroteev}}, \bibinfo {author} {\bibfnamefont {P.~M.}\ \bibnamefont
  {Echenique}}, \ and\ \bibinfo {author} {\bibfnamefont {E.~V.}\ \bibnamefont
  {Chulkov}},\ }\href {http://link.aps.org/doi/10.1103/PhysRevLett.108.246802}
  {\bibfield  {journal} {\bibinfo  {journal} {Physical Review Letters}\
  }\textbf {\bibinfo {volume} {108}},\ \bibinfo {pages} {246802} (\bibinfo
  {year} {2012})}\BibitemShut {NoStop}%
\bibitem [{\citenamefont {Bahramy}\ \emph {et~al.}(2012)\citenamefont
  {Bahramy}, \citenamefont {Yang}, \citenamefont {Arita},\ and\ \citenamefont
  {Nagaosa}}]{bahramy12}%
  \BibitemOpen
  \bibfield  {author} {\bibinfo {author} {\bibfnamefont {M.~S.}\ \bibnamefont
  {Bahramy}}, \bibinfo {author} {\bibfnamefont {B.-J.}\ \bibnamefont {Yang}},
  \bibinfo {author} {\bibfnamefont {R.}~\bibnamefont {Arita}}, \ and\ \bibinfo
  {author} {\bibfnamefont {N.}~\bibnamefont {Nagaosa}},\ }\href {\doibase
  10.1038/ncomms1679} {\bibfield  {journal} {\bibinfo  {journal} {Nature
  Communications}\ }\textbf {\bibinfo {volume} {3}},\ \bibinfo {pages} {679}
  (\bibinfo {year} {2012})}\BibitemShut {NoStop}%
\bibitem [{\citenamefont {Xi}\ \emph {et~al.}(2013)\citenamefont {Xi},
  \citenamefont {Ma}, \citenamefont {Liu}, \citenamefont {Chen}, \citenamefont
  {Ku}, \citenamefont {Berger}, \citenamefont {Martin}, \citenamefont
  {Tanner},\ and\ \citenamefont {Carr}}]{xi13}%
  \BibitemOpen
  \bibfield  {author} {\bibinfo {author} {\bibfnamefont {X.}~\bibnamefont
  {Xi}}, \bibinfo {author} {\bibfnamefont {C.}~\bibnamefont {Ma}}, \bibinfo
  {author} {\bibfnamefont {Z.}~\bibnamefont {Liu}}, \bibinfo {author}
  {\bibfnamefont {Z.}~\bibnamefont {Chen}}, \bibinfo {author} {\bibfnamefont
  {W.}~\bibnamefont {Ku}}, \bibinfo {author} {\bibfnamefont {H.}~\bibnamefont
  {Berger}}, \bibinfo {author} {\bibfnamefont {C.}~\bibnamefont {Martin}},
  \bibinfo {author} {\bibfnamefont {D.~B.}\ \bibnamefont {Tanner}}, \ and\
  \bibinfo {author} {\bibfnamefont {G.~L.}\ \bibnamefont {Carr}},\ }\href
  {http://link.aps.org/doi/10.1103/PhysRevLett.111.155701} {\bibfield
  {journal} {\bibinfo  {journal} {Physical Review Letters}\ }\textbf {\bibinfo
  {volume} {111}},\ \bibinfo {pages} {155701} (\bibinfo {year}
  {2013})}\BibitemShut {NoStop}%
\bibitem [{\citenamefont {Tran}\ \emph {et~al.}(2014)\citenamefont {Tran},
  \citenamefont {Levallois}, \citenamefont {Lerch}, \citenamefont {Teyssier},
  \citenamefont {Kuzmenko}, \citenamefont {Aut{\`e}s}, \citenamefont {Yazyev},
  \citenamefont {Ubaldini}, \citenamefont {Giannini}, \citenamefont {van~der
  Marel},\ and\ \citenamefont {Akrap}}]{tran14}%
  \BibitemOpen
  \bibfield  {author} {\bibinfo {author} {\bibfnamefont {M.~K.}\ \bibnamefont
  {Tran}}, \bibinfo {author} {\bibfnamefont {J.}~\bibnamefont {Levallois}},
  \bibinfo {author} {\bibfnamefont {P.}~\bibnamefont {Lerch}}, \bibinfo
  {author} {\bibfnamefont {J.}~\bibnamefont {Teyssier}}, \bibinfo {author}
  {\bibfnamefont {A.~B.}\ \bibnamefont {Kuzmenko}}, \bibinfo {author}
  {\bibfnamefont {G.}~\bibnamefont {Aut{\`e}s}}, \bibinfo {author}
  {\bibfnamefont {O.~V.}\ \bibnamefont {Yazyev}}, \bibinfo {author}
  {\bibfnamefont {A.}~\bibnamefont {Ubaldini}}, \bibinfo {author}
  {\bibfnamefont {E.}~\bibnamefont {Giannini}}, \bibinfo {author}
  {\bibfnamefont {D.}~\bibnamefont {van~der Marel}}, \ and\ \bibinfo {author}
  {\bibfnamefont {A.}~\bibnamefont {Akrap}},\ }\href
  {http://link.aps.org/doi/10.1103/PhysRevLett.112.047402} {\bibfield
  {journal} {\bibinfo  {journal} {Physical Review Letters}\ }\textbf {\bibinfo
  {volume} {112}},\ \bibinfo {pages} {047402} (\bibinfo {year}
  {2014})}\BibitemShut {NoStop}%
\bibitem [{\citenamefont {Chen}\ \emph {et~al.}(2013)\citenamefont {Chen},
  \citenamefont {Kanou}, \citenamefont {Liu}, \citenamefont {Zhang},
  \citenamefont {Sobota}, \citenamefont {Leuenberger}, \citenamefont {Mo},
  \citenamefont {Zhou}, \citenamefont {Yang}, \citenamefont {Kirchmann},
  \citenamefont {Lu}, \citenamefont {Moore}, \citenamefont {Hussain},
  \citenamefont {Shen}, \citenamefont {Qi},\ and\ \citenamefont
  {Sasagawa}}]{chen13}%
  \BibitemOpen
  \bibfield  {author} {\bibinfo {author} {\bibfnamefont {Y.~L.}\ \bibnamefont
  {Chen}}, \bibinfo {author} {\bibfnamefont {M.}~\bibnamefont {Kanou}},
  \bibinfo {author} {\bibfnamefont {Z.~K.}\ \bibnamefont {Liu}}, \bibinfo
  {author} {\bibfnamefont {H.~J.}\ \bibnamefont {Zhang}}, \bibinfo {author}
  {\bibfnamefont {J.~A.}\ \bibnamefont {Sobota}}, \bibinfo {author}
  {\bibfnamefont {D.}~\bibnamefont {Leuenberger}}, \bibinfo {author}
  {\bibfnamefont {S.~K.}\ \bibnamefont {Mo}}, \bibinfo {author} {\bibfnamefont
  {B.}~\bibnamefont {Zhou}}, \bibinfo {author} {\bibfnamefont {S.-L.}\
  \bibnamefont {Yang}}, \bibinfo {author} {\bibfnamefont {P.~S.}\ \bibnamefont
  {Kirchmann}}, \bibinfo {author} {\bibfnamefont {D.~H.}\ \bibnamefont {Lu}},
  \bibinfo {author} {\bibfnamefont {R.~G.}\ \bibnamefont {Moore}}, \bibinfo
  {author} {\bibfnamefont {Z.}~\bibnamefont {Hussain}}, \bibinfo {author}
  {\bibfnamefont {Z.~X.}\ \bibnamefont {Shen}}, \bibinfo {author}
  {\bibfnamefont {X.~L.}\ \bibnamefont {Qi}}, \ and\ \bibinfo {author}
  {\bibfnamefont {T.}~\bibnamefont {Sasagawa}},\ }\href
  {http://dx.doi.org/10.1038/nphys2768} {\bibfield  {journal} {\bibinfo
  {journal} {Nat Phys}\ }\textbf {\bibinfo {volume} {9}},\ \bibinfo {pages}
  {704} (\bibinfo {year} {2013})}\BibitemShut {NoStop}%
\bibitem [{\citenamefont {Rusinov}\ \emph {et~al.}(2013)\citenamefont
  {Rusinov}, \citenamefont {Nechaev}, \citenamefont {Eremeev}, \citenamefont
  {Friedrich}, \citenamefont {Bl\"ugel},\ and\ \citenamefont
  {Chulkov}}]{rusinov13}%
  \BibitemOpen
  \bibfield  {author} {\bibinfo {author} {\bibfnamefont {I.~P.}\ \bibnamefont
  {Rusinov}}, \bibinfo {author} {\bibfnamefont {I.~A.}\ \bibnamefont
  {Nechaev}}, \bibinfo {author} {\bibfnamefont {S.~V.}\ \bibnamefont
  {Eremeev}}, \bibinfo {author} {\bibfnamefont {C.}~\bibnamefont {Friedrich}},
  \bibinfo {author} {\bibfnamefont {S.}~\bibnamefont {Bl\"ugel}}, \ and\
  \bibinfo {author} {\bibfnamefont {E.~V.}\ \bibnamefont {Chulkov}},\
  }\href@noop {} {\bibfield  {journal} {\bibinfo  {journal} {Phys. Rev. B}\
  }\textbf {\bibinfo {volume} {87}},\ \bibinfo {pages} {205103} (\bibinfo
  {year} {2013})}\BibitemShut {NoStop}%
\bibitem [{\citenamefont {Jacimovic}\ \emph {et~al.}(2014)\citenamefont
  {Jacimovic}, \citenamefont {Mettan}, \citenamefont {Pisoni}, \citenamefont
  {Gaal}, \citenamefont {Katrych}, \citenamefont {Demko}, \citenamefont
  {Akrap}, \citenamefont {Forro}, \citenamefont {Berger}, \citenamefont
  {Bugnon},\ and\ \citenamefont {Magrez}}]{jacimovic14}%
  \BibitemOpen
  \bibfield  {author} {\bibinfo {author} {\bibfnamefont {J.}~\bibnamefont
  {Jacimovic}}, \bibinfo {author} {\bibfnamefont {X.}~\bibnamefont {Mettan}},
  \bibinfo {author} {\bibfnamefont {A.}~\bibnamefont {Pisoni}}, \bibinfo
  {author} {\bibfnamefont {R.}~\bibnamefont {Gaal}}, \bibinfo {author}
  {\bibfnamefont {S.}~\bibnamefont {Katrych}}, \bibinfo {author} {\bibfnamefont
  {L.}~\bibnamefont {Demko}}, \bibinfo {author} {\bibfnamefont
  {A.}~\bibnamefont {Akrap}}, \bibinfo {author} {\bibfnamefont
  {L.}~\bibnamefont {Forro}}, \bibinfo {author} {\bibfnamefont
  {H.}~\bibnamefont {Berger}}, \bibinfo {author} {\bibfnamefont
  {P.}~\bibnamefont {Bugnon}}, \ and\ \bibinfo {author} {\bibfnamefont
  {A.}~\bibnamefont {Magrez}},\ }\href {\doibase
  http://dx.doi.org/10.1016/j.scriptamat.2013.12.017} {\bibfield  {journal}
  {\bibinfo  {journal} {Scripta Materialia}\ }\textbf {\bibinfo {volume}
  {76}},\ \bibinfo {pages} {69} (\bibinfo {year} {2014})}\BibitemShut {NoStop}%
\bibitem [{\citenamefont {Dressel}\ and\ \citenamefont
  {Gr{\"u}ner}(2001)}]{dressel-book}%
  \BibitemOpen
  \bibfield  {author} {\bibinfo {author} {\bibfnamefont {M.}~\bibnamefont
  {Dressel}}\ and\ \bibinfo {author} {\bibfnamefont {G.}~\bibnamefont
  {Gr{\"u}ner}},\ }\href@noop {} {\emph {\bibinfo {title} {Electrodynamics of
  Solids}}}\ (\bibinfo  {publisher} {Cambridge University Press},\ \bibinfo
  {address} {Cambridge},\ \bibinfo {year} {2001})\BibitemShut {NoStop}%
\bibitem [{\citenamefont {Kuzmenko}(2005)}]{kuzmenko05}%
  \BibitemOpen
  \bibfield  {author} {\bibinfo {author} {\bibfnamefont {A.~B.}\ \bibnamefont
  {Kuzmenko}},\ }\href
  {http://scitation.aip.org/content/aip/journal/rsi/76/8/10.1063/1.1979470}
  {\bibfield  {journal} {\bibinfo  {journal} {Review of Scientific
  Instruments}\ }\textbf {\bibinfo {volume} {76}},\  (\bibinfo {year}
  {2005})}\BibitemShut {NoStop}%
\bibitem [{\citenamefont {Kul'bachinskii}\ \emph {et~al.}(2010)\citenamefont
  {Kul'bachinskii}, \citenamefont {Kytin}, \citenamefont {Lavrukhina},
  \citenamefont {Kuznetsov},\ and\ \citenamefont
  {Shevelkov}}]{kulbachinskii10}%
  \BibitemOpen
  \bibfield  {author} {\bibinfo {author} {\bibfnamefont {V.}~\bibnamefont
  {Kul'bachinskii}}, \bibinfo {author} {\bibfnamefont {V.}~\bibnamefont
  {Kytin}}, \bibinfo {author} {\bibfnamefont {Z.}~\bibnamefont {Lavrukhina}},
  \bibinfo {author} {\bibfnamefont {A.}~\bibnamefont {Kuznetsov}}, \ and\
  \bibinfo {author} {\bibfnamefont {A.}~\bibnamefont {Shevelkov}},\ }\href@noop
  {} {\bibfield  {journal} {\bibinfo  {journal} {Electrical and optical
  properties of semiconductors}\ }\textbf {\bibinfo {volume} {44}},\ \bibinfo
  {pages} {1596} (\bibinfo {year} {2010})}\BibitemShut {NoStop}%
\bibitem [{\citenamefont {Akrap}\ and\ \citenamefont
  {et~al.}()}]{KerrUnpublished}%
  \BibitemOpen
  \bibfield  {author} {\bibinfo {author} {\bibfnamefont {A.}~\bibnamefont
  {Akrap}}\ and\ \bibinfo {author} {\bibnamefont {et~al.}},\ }\href@noop {}
  {}\bibinfo {note} {Unpublished}\BibitemShut {NoStop}%
\bibitem [{\citenamefont {Puga}\ \emph {et~al.}(1974)\citenamefont {Puga},
  \citenamefont {Borets},\ and\ \citenamefont {Chepur}}]{puga74}%
  \BibitemOpen
  \bibfield  {author} {\bibinfo {author} {\bibfnamefont {G.}~\bibnamefont
  {Puga}}, \bibinfo {author} {\bibfnamefont {A.}~\bibnamefont {Borets}}, \ and\
  \bibinfo {author} {\bibfnamefont {D.}~\bibnamefont {Chepur}},\ }\href@noop {}
  {\bibfield  {journal} {\bibinfo  {journal} {Soviet Physics Semiconductors}\
  }\textbf {\bibinfo {volume} {8}},\ \bibinfo {pages} {748} (\bibinfo {year}
  {1974})}\BibitemShut {NoStop}%
\bibitem [{\citenamefont {Sklyadneva}\ \emph {et~al.}(2012)\citenamefont
  {Sklyadneva}, \citenamefont {Heid}, \citenamefont {Bohnen}, \citenamefont
  {Chis}, \citenamefont {Volodin}, \citenamefont {Kokh}, \citenamefont
  {Tereshchenko}, \citenamefont {Echenique},\ and\ \citenamefont
  {Chulkov}}]{sklyadneva12}%
  \BibitemOpen
  \bibfield  {author} {\bibinfo {author} {\bibfnamefont {I.~Y.}\ \bibnamefont
  {Sklyadneva}}, \bibinfo {author} {\bibfnamefont {R.}~\bibnamefont {Heid}},
  \bibinfo {author} {\bibfnamefont {K.~P.}\ \bibnamefont {Bohnen}}, \bibinfo
  {author} {\bibfnamefont {V.}~\bibnamefont {Chis}}, \bibinfo {author}
  {\bibfnamefont {V.~A.}\ \bibnamefont {Volodin}}, \bibinfo {author}
  {\bibfnamefont {K.~A.}\ \bibnamefont {Kokh}}, \bibinfo {author}
  {\bibfnamefont {O.~E.}\ \bibnamefont {Tereshchenko}}, \bibinfo {author}
  {\bibfnamefont {P.~M.}\ \bibnamefont {Echenique}}, \ and\ \bibinfo {author}
  {\bibfnamefont {E.~V.}\ \bibnamefont {Chulkov}},\ }\href
  {http://link.aps.org/doi/10.1103/PhysRevB.86.094302} {\bibfield  {journal}
  {\bibinfo  {journal} {Physical Review B}\ }\textbf {\bibinfo {volume} {86}},\
  \bibinfo {pages} {094302} (\bibinfo {year} {2012})}\BibitemShut {NoStop}%
\bibitem [{\citenamefont {Damascelli}\ \emph {et~al.}(1997)\citenamefont
  {Damascelli}, \citenamefont {Schulte}, \citenamefont {van~der Marel},\ and\
  \citenamefont {Menovsky}}]{damascelli97}%
  \BibitemOpen
  \bibfield  {author} {\bibinfo {author} {\bibfnamefont {A.}~\bibnamefont
  {Damascelli}}, \bibinfo {author} {\bibfnamefont {K.}~\bibnamefont {Schulte}},
  \bibinfo {author} {\bibfnamefont {D.}~\bibnamefont {van~der Marel}}, \ and\
  \bibinfo {author} {\bibfnamefont {A.~A.}\ \bibnamefont {Menovsky}},\ }\href
  {http://link.aps.org/doi/10.1103/PhysRevB.55.R4863} {\bibfield  {journal}
  {\bibinfo  {journal} {Physical Review B}\ }\textbf {\bibinfo {volume} {55}},\
  \bibinfo {pages} {R4863} (\bibinfo {year} {1997})}\BibitemShut {NoStop}%
\end{thebibliography}%

\end{document}